# Optimizing control of HIV in Kenya

**Brian G. Williams**

South African Centre for Epidemiological Modelling and Analysis (SACEMA), Stellenbosch, South Africa

Correspondence to BrianGerardWilliams@gmail.com

## Abstract

The Joint United Nations Programme on HIV and AIDS (UNAIDS) has embraced an ambitious global target for the implementation of treatment for people living with HIV. This *90-90-90* target would mean that, by 2020, 90% of all those living with HIV should know their status, 90% of these would be on treatment and 90% of these would have fully suppressed plasma viral loads. To reach this target in the next five years presents a major logistical challenge. However, the prevalence of HIV varies greatly by risk groups, age, gender, geography and social conditions. For reasons of equity no-one should be denied access to life-saving anti-retroviral therapy (ART) but for reasons of effectiveness and impact the focus must first be on those who are most likely to be infected with HIV and therefore most likely to infect others.

In Kenya the prevalence of HIV in adults varies by two orders of magnitude among the counties. The effective implementation of *90-90-90* will depend on first providing ART where the prevalence of infection is greatest, then to those that are most easily reached in large numbers and finally to the whole population. Here we use routine data from ante-natal clinics and national survey data to assess the variation of the prevalence of HIV among counties in Kenya; we suggest reasons for this variation, and estimate the effectiveness of targeting the role out of ART.

The highest prevalence occurs in some, but not all, of the counties bordering Lake Victoria and these are most in need of ART. These districts in Nyanza Province, account for 31% of all cases in Kenya even though they only make up 10% of the population and cover 1.8% of the land-area. The highest concentrations of HIV cases are in Nairobi and Mombasa. These two cities account for a further 18% of all cases in Kenya but only make up 12% of the population and cover 0.1% of the land-area. The logistics of providing ART in these two cities will be relatively straightforward given their small geographical area. Finally we note that Nakuru County, on the main highway from Mombasa to Nairobi and then to Uganda, has the next highest prevalence of HIV after the counties in Nyanza Province and the two major cities.

## Introduction

It has become increasingly clear that when people are infected with HIV they should start anti-retroviral treatment (ART) as soon as possible; any delay in the start of treatment increases a persons risk of dying.[1] Furthermore, with good treatment and adherence an infected persons' viral load falls by a factor of 100 within one month and 10,000 within 12 months[2] rendering them uninfectious to other people. In the year 2000 the International AIDS Society[3] and the Department of Health and Human Services[4] recommended immediate ART for all HIV positive people if their CD4 cell count was below 500/μL[5] and this is now supported by the World Health Organization[6] and others.[7]

The fact that early ART is not only in the best interest of the individual person infected with HIV but also has the potential to stop transmission has led the Joint United Nations Programme on HIV and AIDS (UNAIDS) to call for an end to HIV/AIDS centred on their *90-90-90* target: by 2020, 90% of all those living with HIV should have been tested within the last one year, unless of course they already know that they are infected with HIV, 90% of these should be on treatment and 90% of these should have plasma viral loads below 1,000 copies/mL.

Reaching the *90-90-90* target by 2020 is ambitious. It will require many things to be in place including a regular and assured supply of drugs, good community mobilization and support to counter stigma and discrimination, good counselling and support to ensure high levels of adherence and to promote safe sex, other methods of prevention, good patient monitoring and follow up to identify places where the programme may be failing, and good surveillance tools to know whether the targets are being reached.

Given these many challenges, as well as the need to mobilize funding and human resources, it will be essential to plan and implement the programme as efficiently and as cost effectively as possible. In this paper we use data from Kenya to examine, in particular, the geographical distribution of HIV infection to provide support for planning and implementing what will be one of the most challenging public health disease-control programmes ever attempted.

We start by carefully re-examining the available data on HIV in Kenya, largely derived from measurements of the prevalence of HIV in women attending ante-natal clinics (ANC) for each of the counties of Kenya. Since these are the only data on epidemic trends in Kenya, as in most other countries in Africa, it is essential that these data are reviewed very carefully.

Having fitted trends to the available data we consider the counties in three groups: Nyanza Province,* where the prevalence is highest, Nairobi and Mombasa, where the density of infected people is greatest, and the remaining counties, where both the prevalence of infection and the density of infected people is relatively low.

Finally, we consider some explanations for the geographical diversity of the epidemic in Kenya and discuss the implications for achieving the *90-90-90* target by 2020.

* Nyanza Province contains Siaya, Kisumu, Homa Bay and Migori Counties

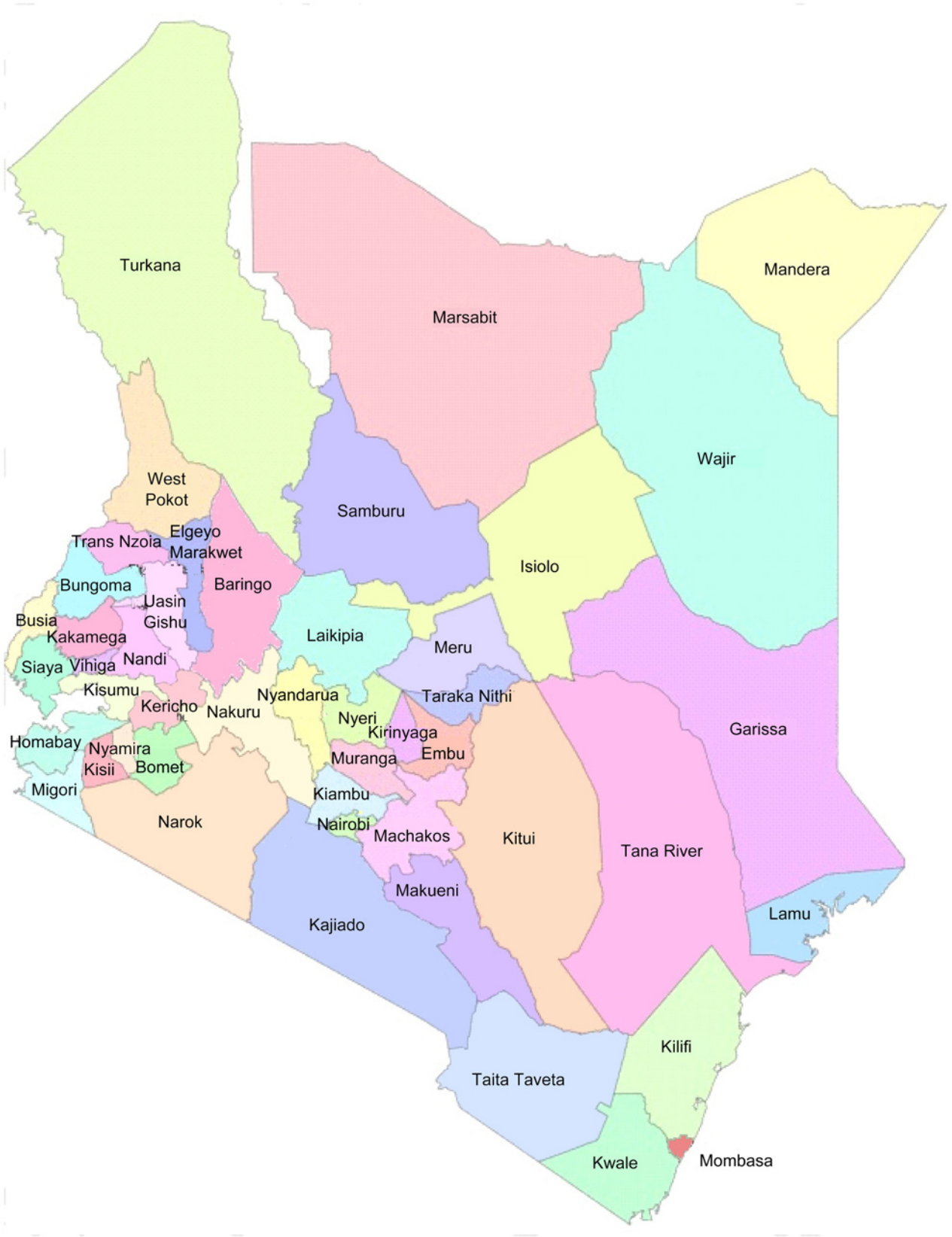

Figure 1. Kenya County names as of 2014.

## Methods

Figure 1 is a map of Kenya with the counties indicated for the purposes of the discussion that follows. We draw on two sets of data. The first gives the prevalence of HIV measured at different times in ANC clinics in each of the counties. These data are available from the 'Spectrum' files held by UNAIDS in Geneva.[8] Unfortunately, these data were not collected consistently over time and according to rigorous standards, and there is considerable variation over and above the binomial errors which needs to be included. Fortunately, a large survey was carried out in 2006 in which 564,253 women attending ante-natal clinics were tested for HIV from all counties in Kenya[9] and this provides a key marker that determines the prevalence of infection in each county. The geographical distribution of the prevalence of HIV in 2006 is given in Figure 2 where the upper part of the figure is for the whole country and the lower part shows a detailed distribution in Nyanza Province and the adjacent districts.

In all counties of Kenya the prevalence rose to a peak, fell significantly and then levelled off to an asymptote. To fit the data we therefore proceed as follows. We fit a double-logistic function to the available trend data for each county in Kenya so that the prevalence of infection at time $t$ is

$$P(t) = a_1 \frac{e^{r_1(t-t_1)}}{1-e^{r_1(t-t_1)}} - a_2\left(1-\frac{e^{r_2(t-t_2)}}{1-e^{r_2(t-t_2)}}\right) \qquad 1$$

The first term in Equation 1 accounts for the rate, $r_1$, the timing, $t_1$, and the peak, $a_1$, of the initial rise in the prevalence. The second term accounts for the rate, $r_2$, the timing $t_2$, and the subsequent drop, $a_2$, in the prevalence as a result of reductions in transmission not apparently associated with the increasing coverage of ART. The reasons for, and drivers of, these reductions in transmission are not understood and we discuss this further below.

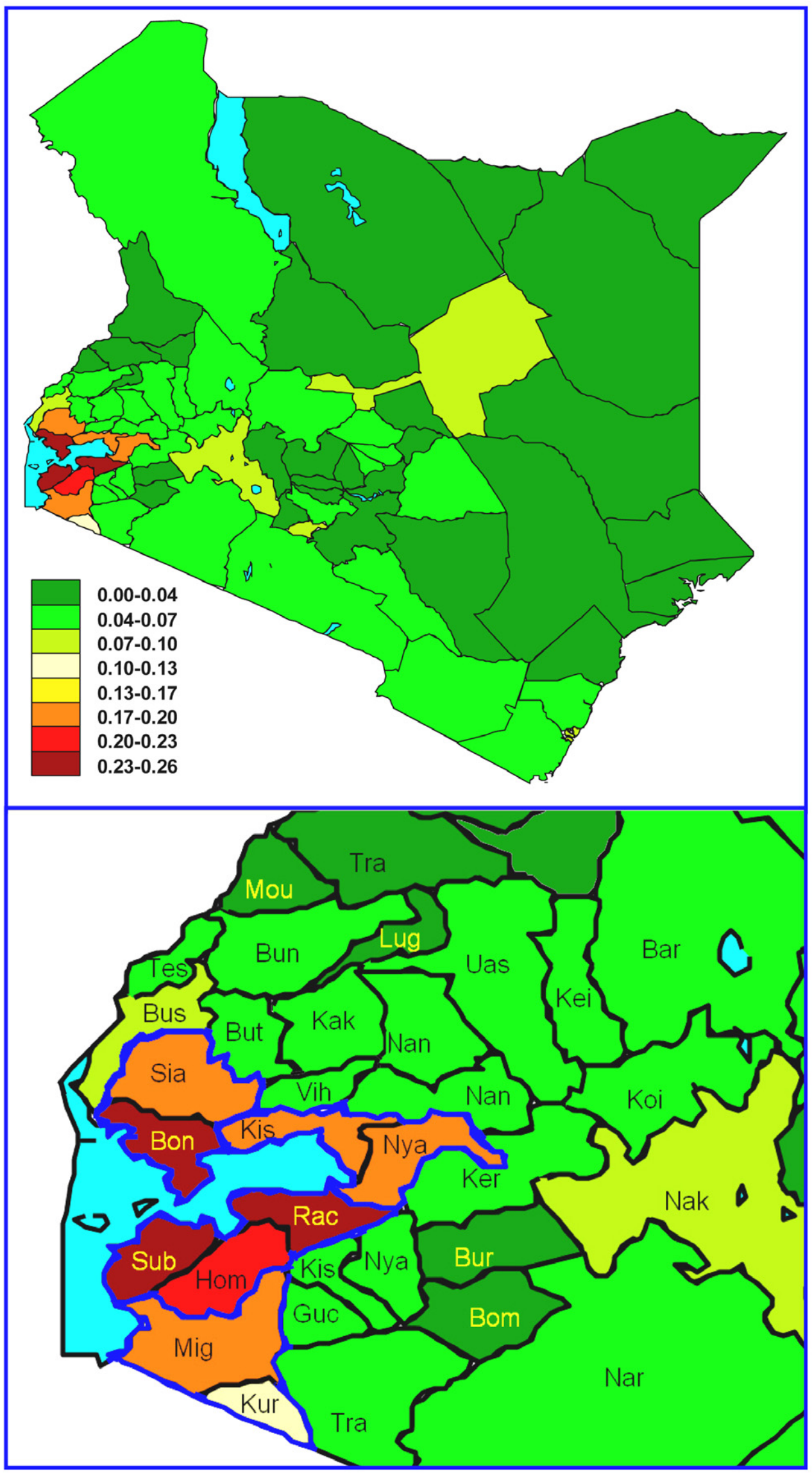

Figure 2. HIV prevalence among women attending ante-natal clinics in each district of Kenya in 2006. County names are given in Figure 1 and abbreviated here. Bondo District is now included in Siaya County, Nyanza District in Kisumu County, Rachuonyo and Suba districts in Homa Bay County, and Kuria district in Migori County.

We start with a conventional least squares fit to those counties with reasonably consistent data but with the proviso that the data points were scaled, in each county, by a factor, $\sigma$, chosen to ensure that the fitted curve passes exactly through the measured prevalence for that county in 2006. This factor $\sigma$ was allowed to vary in the optimization routine and is discussed further below.

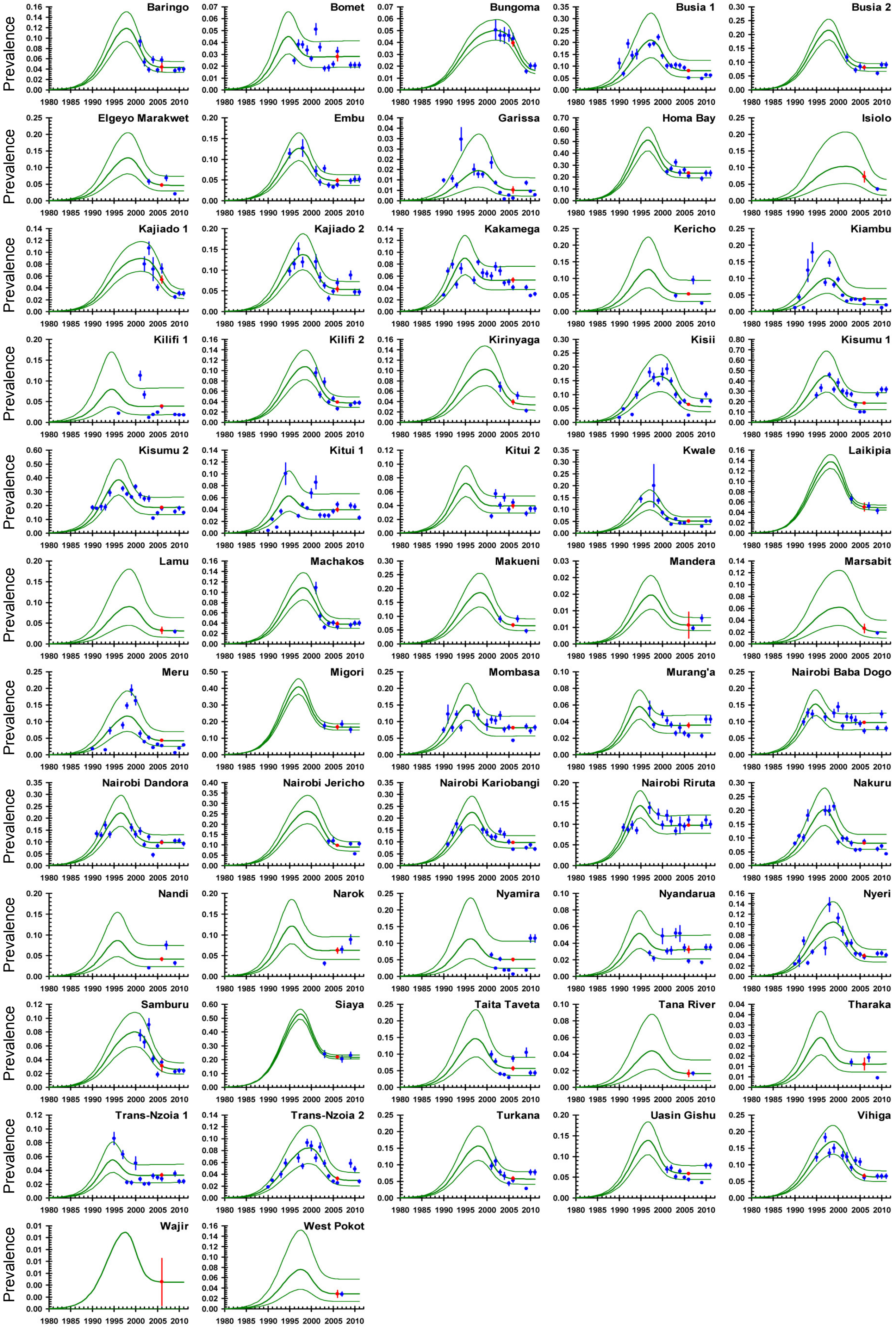

Figure 3. HIV prevalence in ante-natal clinics in Kenya between 1980 and 2012. Blue dots: routine surveys. Red dots: 2006 survey. Green lines: Fitted curves, all with 95% confidence limits.

For each of the parameters in Equation 1 we then calculate the mean and the standard deviation across all of the counties for which the data are reasonably consistent over several years and use the distributions of these parameters and the constraint on the scale factor, $\sigma$, as prior estimates in a Bayesian fit to the data. In Nyanza Province the peak prevalence greatly exceeded that in any other province, and in Nairobi and Mombassa, the epidemic started earlier than in other counties. We therefore used three different sets of prior distributions for Nyanza, Nairobi and Mombassa, and for the remaining counties.

When combining the data across districts we weighted the prevalence in each district by the 2014 estimates of the adult population in each district[10] to calculate the overall prevalence.

## Results

The 2006 survey is a critical reference year for the prevalence of HIV in Kenya.[9] It shows clearly that in 2006 the ANC prevalence of HIV varied from about 2% to 5% in most counties but ranged from 17% to 26% in Nyanza Province. It is interesting to note (Figure 2) that in Busia, which borders both Uganda and Lake Victoria, the prevalence of HIV was only 8%, about one-third of the prevalence in the other counties neighbouring Lake Victoria. We also see in Figure 2 that the prevalence in Mombasa, Nairobi and Nakuru was relatively high, at about 15%, and note that these three towns are on the main highway from Mombasa to Uganda.

The fitted data, with 95% confidence limits on the data points and the fitted curves are given in Figure 3. The blue dots are the routine ANC data and one sees that they are very variable. The data for Busia, Embu, Nakuru and Trans Nzoia are reasonably consistent while the data for Isiolo, Lamu, Mandera, Narok and several other counties are limited. Nevertheless the combination of the ANC data, the results of the very extensive survey in 2006, and the use of Bayesian priors give reasonably convincing fits across the whole country (Figure 3). We also compare the distributions of the key parameters from the fits among the counties to ensure consistency and to understand differences in the epidemic in different counties.

For all three groups of counties the mean value of the scale parameter in the Appendix, Figure 6 is close to 1 which is important since it suggests that, on average, the estimates from the 2006 survey are in agreement with the routine ANC data. There is, however, some variation about the mean, especially in the low prevalence group of countries. In Laikipia, for example, the routine ANC data have to be almost doubled to bring them up to the value obtained in the 2006 survey, while in Garissa and Taraka Nithi the routine data have to be reduced by a factor of 5.

In all three groups of counties the data in Figure 7 suggest that the prevalence of HIV reached half the maximum value in about 1993 so that it seems that the epidemic took off at about the same time in all areas of the country but may have risen slightly earlier in Nairobi and Mombasa than in the rest of the country. One should note that in many of the low prevalence counties there was very little data to suggest when the initial rise occurred. In Figure 7 the Bayesian priors mean that a very similar value for the initial rise is obtained across all counties for which the early data are weak or absent.

The timing of the fall in prevalence (Appendix, Figure 8), is also consistent across the three groups of counties and the decline reached half its value in 2000 although there is some variation among counties with some falling to half the final asymptote in 1996 and others only in 2006. On average it seems that the fall occurred about 7 years after the rise to the peak.

The initial rates of increase, 0.4/year (Figure 9), and decline, 0.9/year (Figure 10), are also similar across all counties but this also reflects the fact that few counties have good data on the initial rise in the prevalence and the Bayesian prior largely determines the outcome where the data are weak. Nevertheless it is interesting to note that the prevalence converges to the eventual asymptote at about half of the rate at which the prevalence rises at the start of the epidemic.

The peak value of the prevalence of infection (Figure 4and Appendix Figure 11) varies substantially across the three groups of countries. In Nyanza Province, Homa Bay, Kisumu, Migori and Siaya, the average value of the peak prevalence was 40%, in Nairobi and Mombasa it was less than half of this at 18%, and in the rest of the country it was only 8%. This variation demands an explanation and has implications for how to roll out ART across the country.

The decline from the peak is consistent across the three groups of counties, Figure 12. In all three the prevalence falls to about 40% of its peak value but then stabilizes at lower steady state.

From these fits we obtain the weighted prevalence of infection in the three groups of counties (Figure 4). This shows clearly that the timing and initial rate of increase and the timing and the extent of the decline from the peak are similar in the three groups of counties but the peak occurred about one year earlier in Nairobi and Mombasa than in Nyanza Province.

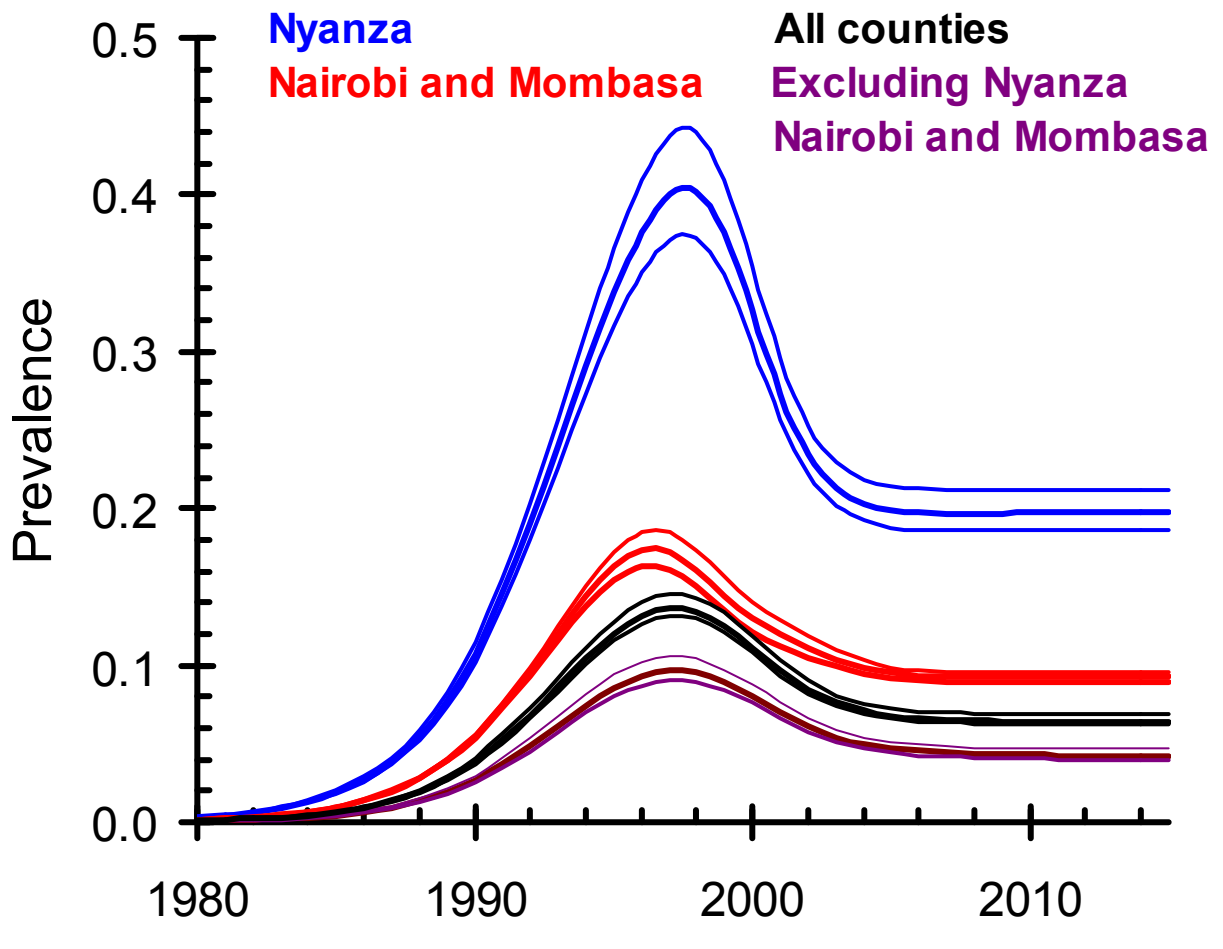

Figure 4. The prevalence of HIV in different parts of Kenya. Blue: Nyanza Province; Red: Nairobi and Mombasa; Black: all of Kenya; Purple: all excluding the red and blue areas.

## Discussion

The data and this analysis raises a number of important questions concerning the natural history of HIV in Kenya but the trend data are limited and not always consistent over time and the 2006 survey is particularly important as it enables us to determine the absolute value of the prevalence with reasonable confidence in all counties.

The most important observation is the very high prevalence of infection in Nyanza Province even as compared to the immediately adjacent counties, especially Busia. It is known that the people of Nyanza Province do not carry out male circumcision by tradition and this is, of course, the reason why studies of male circumcision in Kenya have been carried out there.[11] While it is clear that male circumcision reduces a man's risk of infection by about 60% one might expect this to have a larger effect on the initial rate of increase than on the peak prevalence; the peak prevalence is more likely to be determined by the extent of heterogeneity in the risk of infection. In this regard, Busia County is of particular interest. Even though it borders both Lake Victoria and Uganda the prevalence in Busia (Figure 3) is less than half of the prevalence in the four counties of Nyanza (Figure 4). A careful comparison of Busia with Siaya, Kisumu, Homa Bay and Migori could yield important insights into the determinants of HIV in Kenya.

It has been suggested that labour migration is the reason why the nine countries, worst affected by HIV, are all in southern Africa.[12-17] It is clear from the age distribution of men and women in Nairobi and Mombasa that there is considerable migration to both of these cities in people aged 15 to 25 years, with migration of women preceding that of men by about three years of age.

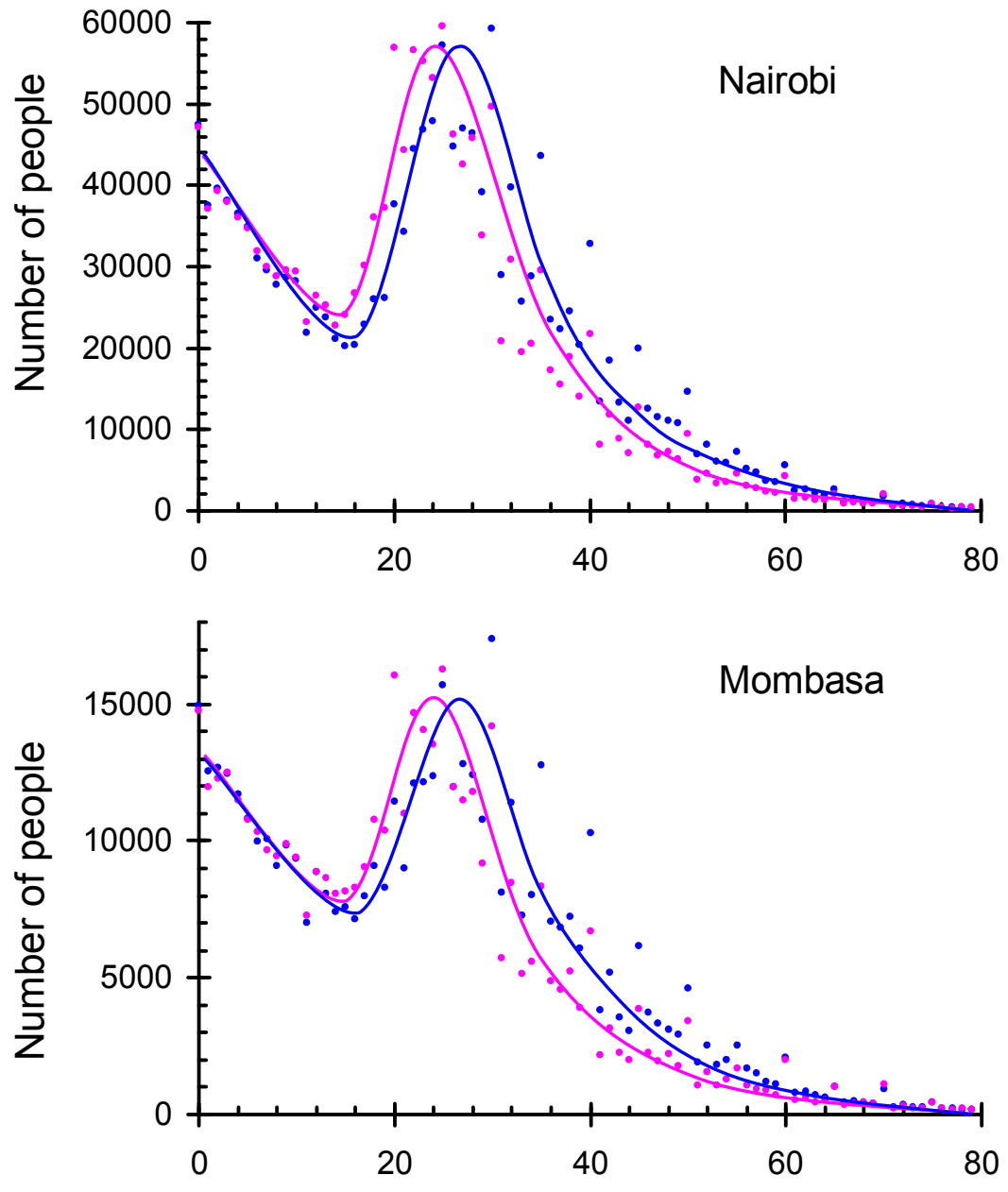

Figure 5. The age and sex distribution of men (blue) and women (pink) in Nairobi and Mombasa in 2012.

It is perhaps not surprising that the prevalence of infection in Nairobi and Mombasa is about three times greater than in the rest of the country, excluding Nyanza Province. In this regard one should note that the main highway linking the outside world to Uganda starts in Mombasa, passes through Nairobi and Nakuru and then through Eldoret on the way to Kampala. The link to the epicentre of the HIV epidemic in East Africa and the movement of truck drivers and other migrant workers along this highway are both likely to be key determinants of the spread of HIV. It is also significant that the prevalence of infection in Nakuru, a major struck-stop 0n the Mombasa-Kampala highway, is similar to the prevalence of infection in Nairobi and Mombasa and higher than in any other county apart from those in Nyanza Province.

From the fits to the prevalence data and the adult population of each county we estimate the number of people living with HIV in Kenya as shown in Table 1.

Table 1. The number of people living with HIV, the total population, and the land area for Nyanza County, Nairobi and Mombasa. Numbers in brackets are percentages of the total.

| | HIV (k) | Pop. (M) | Area (km)$^2$ |
|---|---|---|---|
| Total | 1,363 | 24.4 | 582,650 |
| Nyanza | 416 (31%) | 2.5 (10%) | 10,377 (1.8%) |
| Nairobi & Mombasa | 242 (18%) | 3.0 (12%) | 649 (0.11%) |
| Nyanza, Nairobi & Momb. | 658 (48%) | 4.7 (19%) | 11,026 (1.9%) |

It seems likely that the epidemic of HIV in Kenya spread from Uganda along the Mombasa-Kampala highway. For reasons that are still not certain it then affected people living in Nyanza to a greater extent than people living elsewhere along the highway, and then spread out from the highway to people living in the rest of the country.

The immediate priority is to bring the epidemic under control in Nyanza Province which has 31% of all those living with HIV in Kenya but only includes 10% of the total population and 1.8% of the land area. Nairobi and Mombasa account for a further 18% of all those living with HIV in Kenya and although they make up 12% of the population they are concentrated in 0.11% of the land area. The logistics of delivering drugs and providing services to people living in such highly concentrated areas will be much simpler than providing services in the remaining 98% of the country.

Since Mombasa, Nairobi, Nakuru and Nyanza all lie along the Mombasa-Kampala highway this clearly facilitates the spread of HIV but also facilitates the provision of services.

It is very important that the other counties are not forgotten but the approach to the control of HIV in these counties will have to be different. First of all it is likely that controlling HIV in the high prevalence counties will have an indirect effect on the low prevalence counties. Second, to find people living in often remote and scattered areas might be more efficiently done through contact tracing, especially if people who migrate to Nairobi are identified and their local contacts tested for HIV. The logistics of delivering drugs to these remote places will remain a challenge.

## Conclusion

Reaching the *90-90-90* UNAIDS target for HIV testing, ART treatment and levels of compliance will be challenging. However, those living in Nyanza Province are much more likely to be infected with HIV than people living elsewhere in Kenya and the geographical concentration of people living with HIV is much greater in Nairobi and Mombasa than anywhere else in Kenya. If the *90-90-90* targets can be reached first in Nyanza province and then in Nairobi and Mombasa with additional focus on the trucking route from Mombasa to Kampala, this alone should cover nearly half of all those living with HIV while only needing to access less than one-quarter of the whole population in a geographical area covering only 2% of the country.

Focussing initially on Nyanza Province, Nairobi and Mombasa will make universal testing, drug delivery and community support much easier and more effective. In other areas of the country the scattered nature of the population and the relatively low prevalence of infection will need other approaches. Where the prevalence of HIV and the density of people are both low, contact tracing becomes a more viable way of finding, testing, treating and providing support to people infected with HIV.

The initial rate of increase of the prevalence implies a doubling time of 1.7 years and a case reproduction number, $R_0$, of 4.8[18] so that to eliminate HIV in Kenya will need an 80% reduction in transmission. Reaching the *90-90-90* target by 2020 will reduce $R_0$ by 73% to a value of about 1.3. If the roll-out of ART is supported by effective counselling and support, condom promotion, providing access to pre-exposure prophylaxis to people at high risk, especially to female sex workers, and the continued roll out of male circumcision, especially in Nyanza, this should easily provide the extra 30% reduction in transmission needed to bring $R_0$ below 1 and ensure eventual elimination of the epidemic.

Because the initial rate of increase in the prevalence of HIV seems to be fairly consistent across the three key groups of countries, Nyanza, Nairobi and Mombasa, and the rest of Kenya, while the peak prevalence varies by a factor of five, it seems likely that the high rates of HIV in Nyanza as compared to the other counties reflects a greater heterogeneity in risk behaviour rather than a greater likelihood of individual risk per exposure. The reasons for these patterns of infection remain uncertain but should be explored further.

The next stage of this analysis will be to develop dynamical models of the epidemic in order to further test the ideas and suggestions discussed here but also to project future time trends in the epidemic, allowing for the current and future levels of ART coverage, and to carry out detailed cost benefit analyses which will be needed if the response is to be properly planned.

.

## Appendix. Distributions of the fitted parameter values

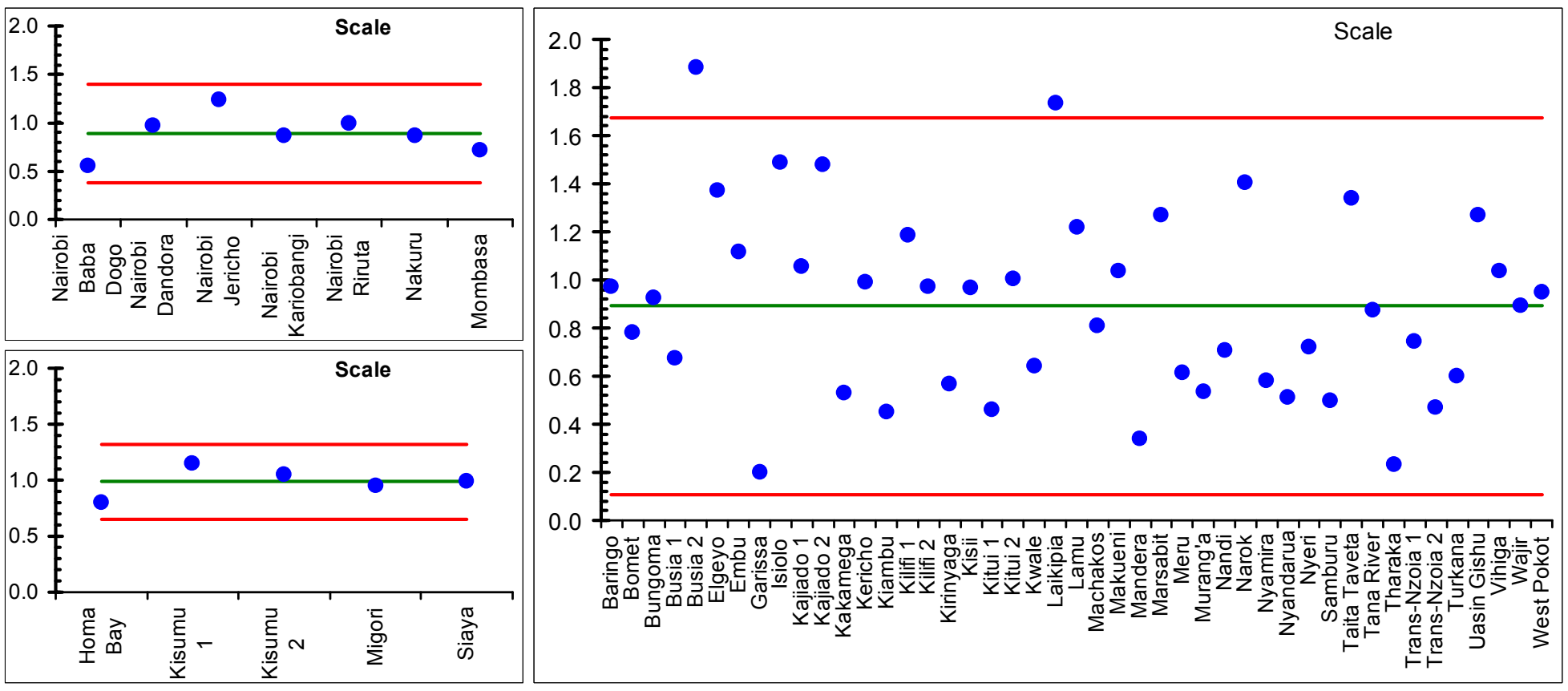

Figure 6 'Scale' is $\sigma$, the adjustment to the routine ANC data so that the fitted values passes through estimate from the 2006.

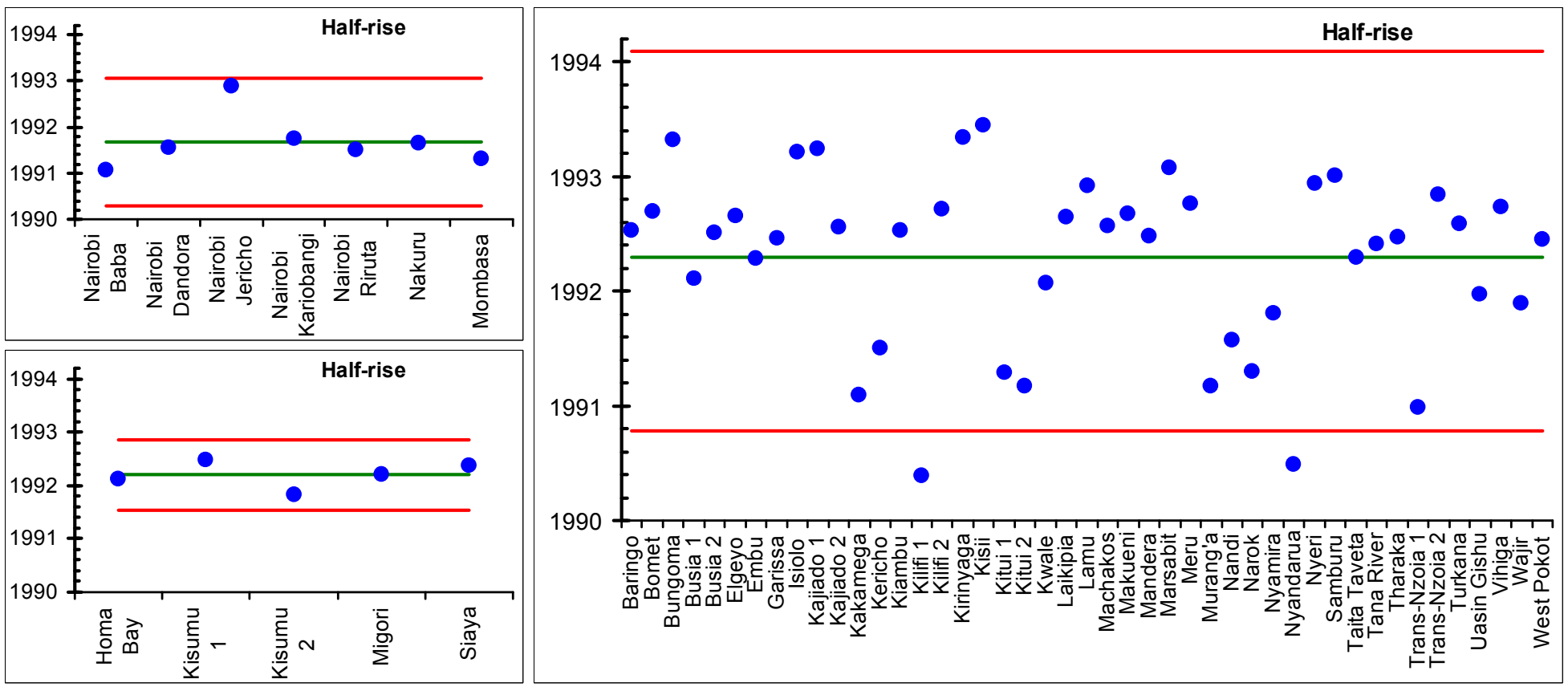

Figure 7. 'Half-rise' is $\tau_1$ the time when the prevalence first reaches half its peak value.

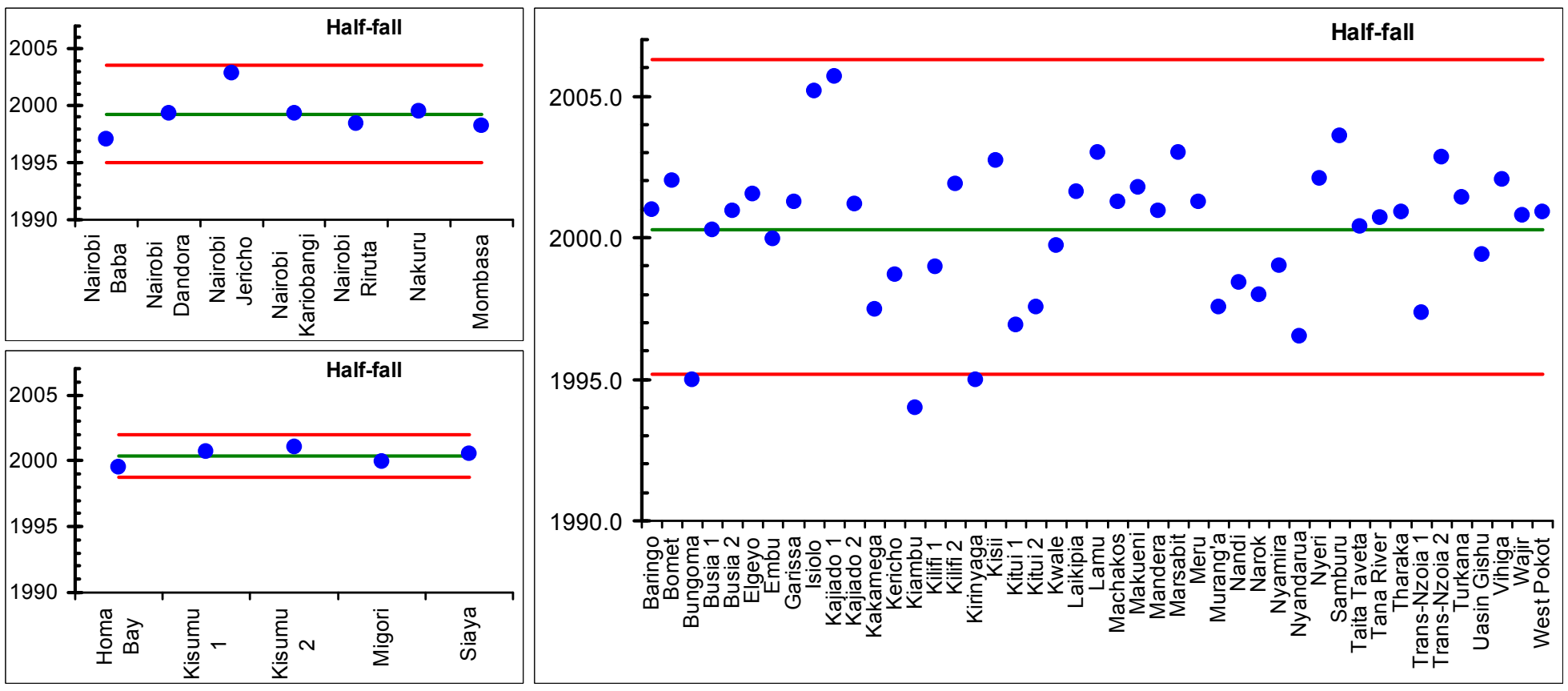

Figure 8. ‘Half-fall’ is $\tau_2$ the time when the prevalence falls to half-way between the peak value and the eventual asymptotic value.

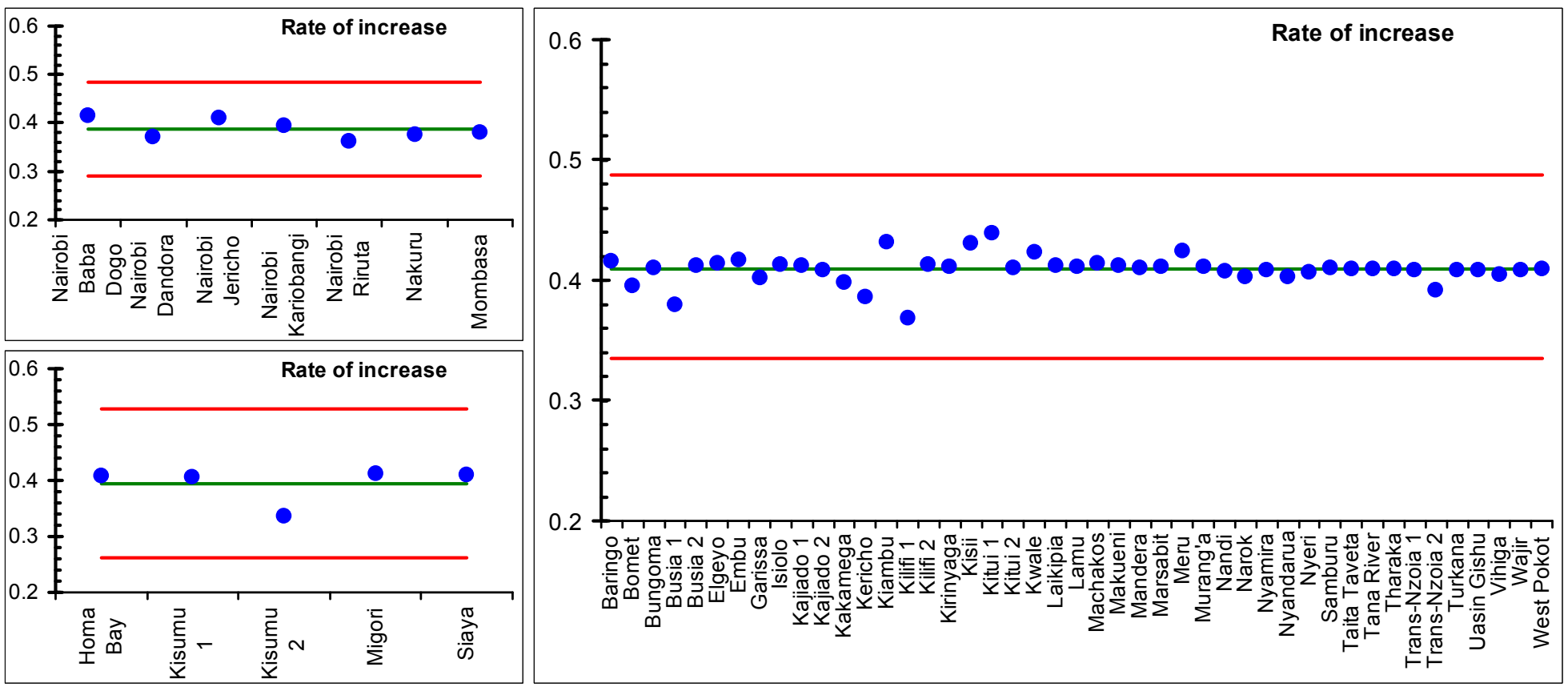

Figure 9. ‘Rate of increase’ is $\rho_1$ the initial rate of increase of the prevalence.

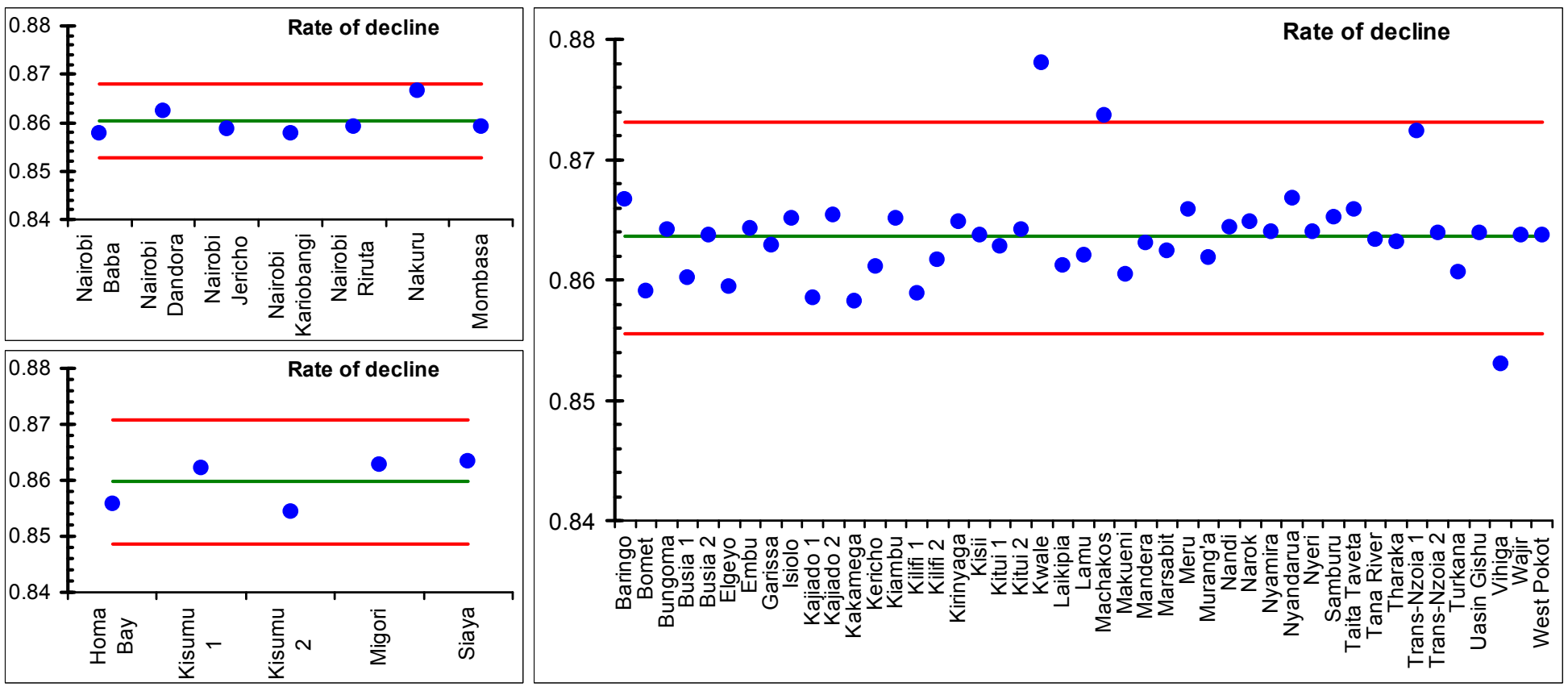

Figure 10. ‘Rate of decline’ is $\rho_2$ the final rate of decline of the prevalence.

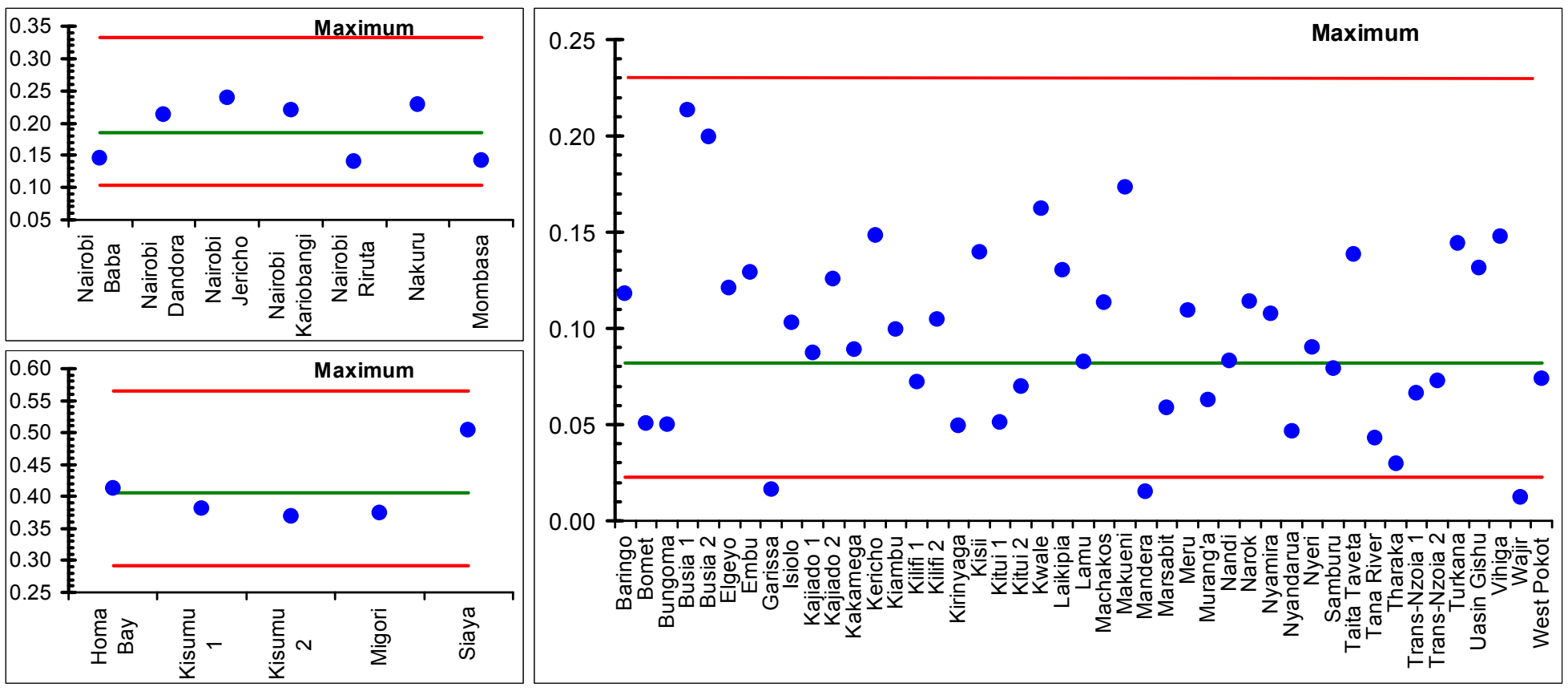

Figure 11. 'Maximum' is the peak value of the prevalence.

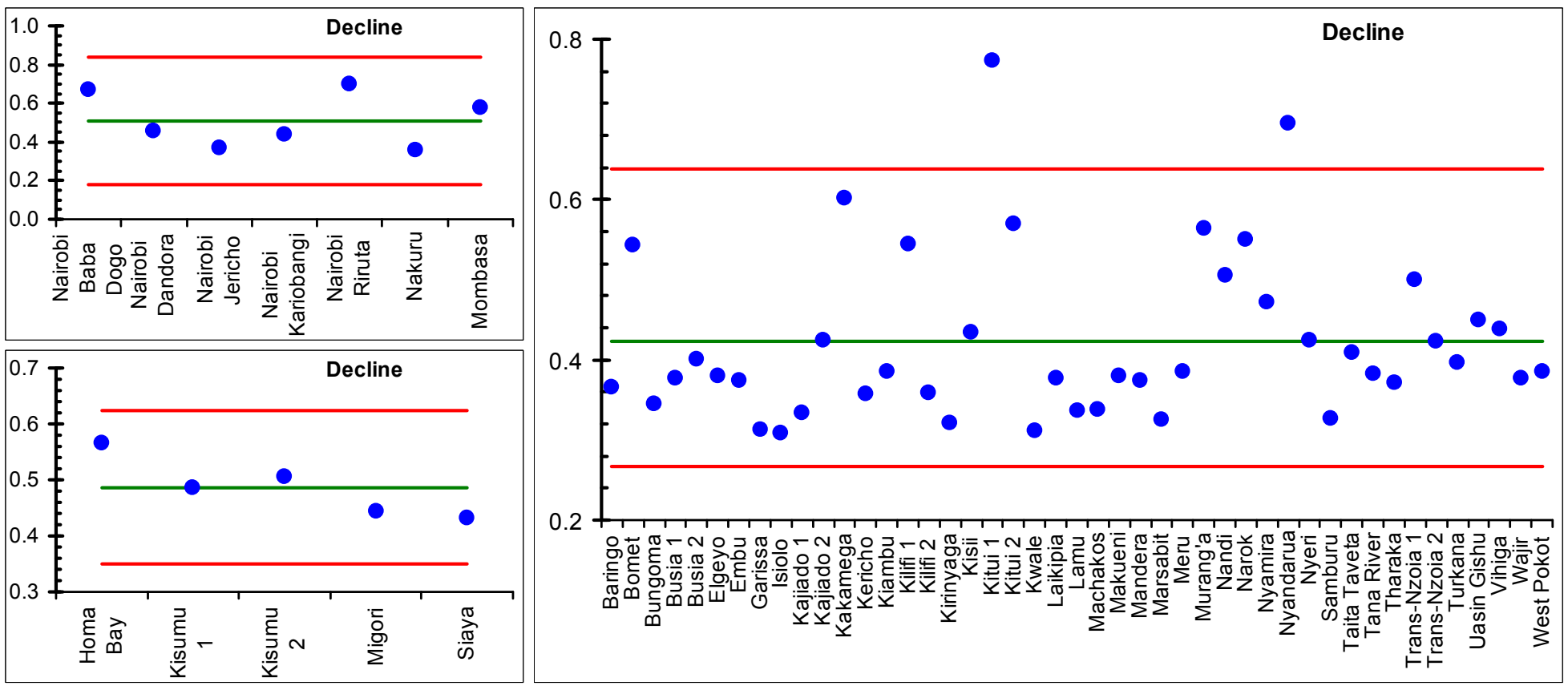

Figure 12. 'Decline' is the ratio of the prevalence in 2012 to the peak value of the prevalence.